# Numerical Investigation of Airborne Infection Risk in an Elevator Cabin under Different Ventilation Designs

Ata Nazari[1], Changchang Wang (王畅畅)[2, 3], Ruichen He (何瑞辰)[4], Farzad Taghizadeh-Hesary[5], Jiarong Hong (洪家荣) [*4, 3]

[1]*University of Tabriz, Department of Mechanical Engineering, Tabriz, Iran*

[2]*Department of Civil and Environmental Engineering, The Hong Kong Polytechnic University, Hung Hom, Kowloon, Hong Kong, China*

[3] *St. Anthony Falls Laboratory, University of Minnesota, Minneapolis, Minnesota 55455, USA*

[4]*Department of Mechanical Engineering and St. Anthony Falls Laboratory, University of Minnesota, Minneapolis, Minnesota 55455, USA*

[5]*ENT and Head and Neck Research Center and Department, The Five Senses Health Institute, School of Medicine, Iran University of Medical Sciences, Tehran, Iran*

## Abstract

Airborne transmission of SARS-CoV-2 via virus-laden aerosols in enclosed spaces poses a significant concern. Elevators, commonly utilized enclosed spaces in modern tall buildings, present a challenge as the impact of varying heating, ventilation, and air conditioning (HVAC) systems on virus transmission within these cabins remains unclear. In this study, we employ computational modeling to examine aerosol transmission within an elevator cabin outfitted with diverse HVAC systems. Using a transport equation, we model aerosol concentration and assess infection risk distribution across passengers' breathing zones. We calculate particle removal efficiency for each HVAC design and introduce a suppression effect criterion to evaluate the effectiveness of the HVAC systems. Our findings reveal that mixing ventilation, featuring both inlet and outlet at the ceiling, proves most efficient in reducing particle spread, achieving a maximum removal efficiency of 79.40% during the exposure time. Conversely, the stratum ventilation model attains a mere removal efficiency of 3.97%. These results underscore the importance of careful HVAC system selection in mitigating the risk of SARS-CoV-2 transmission within elevator cabins.

**Keywords**: Air circulation; COVID-19; Elevator; *OpenFOAM*; SARS-CoV-2; Ventilation

## 1. Introduction

Coronavirus disease 2019 (COVID-19) is primarily transmitted through inhalation of respiratory droplets containing severe acute respiratory syndrome coronavirus 2 (SARS-CoV-2) released during exhalation activities, such as speaking, coughing, or sneezing (Asadi et al., 2020; Lewis, 2020; Wilson et al., 2020; Nazari et al., 2021a; Fadavi et al., 2021; Rakhsha et al., 2020a, b). SARS-CoV-2-laden aerosols, which can range from 0.1 to 100 $\mu m$ in diameter, can remain suspended in the air for several hours and travel long distances, posing a significant infection risk even at social distances beyond 6 feet from an infected individual (Dbouk & Drikakis, 2020; Niazi et al., 2021). The risk of COVID-19 airborne transmission is particularly high in confined indoor spaces, especially in areas with high occupancy rates. Among the various engineering controls, ventilation plays a critical role in reducing airborne infection risk (Bhagat et al., 2020). Guidelines for COVID-19 recommend increasing the ventilation rate to reduce the risk of viral transmission in enclosed spaces (REHVA, 2020; Stewart et al., 2020). Inadequate ventilation systems can result in the circulation of respiratory droplets and aerosols by heating and ventilation air conditioning (HVAC) systems, leading to increased risk of airborne transmission (Bhagat et al., 2020; Avni & Dagan, 2022; Calmet et al., 2021; Dey et al., 2021). Several studies have evaluated SARS-CoV-2 transmission through HVAC systems in various practical settings, including underground car parks (Nazari et al., 2021a), urban subway (Nazari et al., 2022), restaurants (Zhang et al., 2021a; Liu et al., 2021), restrooms (Schreck et al., 2021), elevators (Dbouk & Drikakis, 2021; Nouri et al., 2021; Peng et al., 2021), public spaces (MacIntyre & Hasanain, 2020; Shao et al., 2020), cleanrooms (Rezaei et al., 2020), urban buses (Yang et al., 2022; Zhang et al., 2021c), classrooms (Narayanan et al., 2021; He et al., 2021a; Burgmann et al., 2021), cafeterias (Wu et al., 2021), airplanes (Horstman & Rahai, 2021; Talaat et al., 2021), dental clinics (Komperda et al., 2021; Jia et al., 2021), escalators (Li et al., 2021), orchestral wind instrument performances (Abraham et al., 2021), and other enclosed spaces (Yan & Lan, 2020; Peña-Monferrer et al., 2021; Motamedi et al., 2022; Ahmadzadeh & Shams, 2021; Li et al., 2020; Agrawal and Bhardwaj, 2020). Many studies have investigated airborne transmission and developed corresponding mitigation strategies (Dbouk & Drikakis, 2020; Zhang et al., 2021b; Xi et al., 2020; Koroteeva & ShagiyanKoroteeva & Shagiyanova, 2022; He et al., 2021b; Dbouk & Drikakis, 2020, Dbouk & Drikakis, 2021; Dbouk & Drikakis, 2022; Auvinen et al., 2022; Agrawal & Bhardwaj, 2021).

Elevators are an essential mode of urban transportation, but their small, crowded, and poorly ventilated confined spaces pose a significant risk for COVID-19 transmission. To date, no study has investigated the airborne transmission of COVID-19 in high-speed elevators with different HVAC designs. During the COVID-19 pandemic, this risk was particularly challenging in metropolitan cities with high-rise buildings where people spend more time in elevators, such as New York City. According to the National Elevator Industry, Inc (NEII), people in the United States travel approximately 2.5 billion miles on elevators and escalators each year, which is more than the total rail and air miles traveled (NEII, 2020). Therefore, understanding aerosol transmission and identifying effective and economically viable measures to limit disease transmission in elevator cabins remain pressing matters. The first reported case of COVID-19 transmission inside an elevator cabin in Seoul, South Korea, sparked discussions about elevator safety in communities, especially after COVID-19 reopening (First case of COVID-19 transmission inside an elevator, 2020). The necessity of social distancing has reduced elevators' capacity, leading to lobby queues, long waiting times, and reduced work productivity. Along with social distancing, wearing face masks, and washing hands, improving ventilation in elevators can help reduce viral transmission in confined spaces (Ananthanarayanan et al., 2022). Despite the overwhelming evidence on smoke control during a fire in elevator shafts in high-rise buildings, only a few studies have evaluated the risk of airborne transmission inside elevator cabins. A recent survey on aerosol concentration measurement in public spaces, including elevator cabins, showed that elevators with poor ventilation, defined as having less than 6 air changes per hour (ACH), have a higher risk of infection (Somsen et al., 2020). An experimental study investigated measures to reduce aerosol persistence in hospital elevators by mimicking a single cough in medium- and large-sized elevators with closed doors, open doors, or during operation (van Rijn et al., 2020). The study highlighted the significance of ventilation design in reducing airborne infection risk. Therefore, there is a dire need to systematically investigate different ventilation designs in elevator cabins to minimize the risk of infection while potentially increasing elevator capacity.

During the COVID-19 pandemic, numerous studies have been conducted to assess the risk of infection and the effectiveness of different risk mitigation strategies in various indoor environments using computational fluid dynamics (CFD) tools. These studies have shown the importance of airflow patterns on aerosol transmission. Nazari et al. performed a numerical study to detect the effects of jet fans on spreading respiratory aerosols inside underground car parks. The investigators identified four different locations as sneeze sources inside underground car parks and suggested that safe pathways should be close to the fresh air ducts, with ultraviolet light emitters and high-efficiency particulate air (HEPA) filters inside the jet fans potentially further eliminating the virus (Nazari et al., 2021b). In another study, the same group investigated aerosol spreading inside the urban subway by considering two HVAC systems and various infected person configurations. They proposed a novel HVAC system design by changing the position of exhaust ducts from the urban subway cabin ceiling to near the floor. For the case of the infected individual breathing near the fresh-air ducts, they found that the conventional HVAC system design led to the largest spread and highest aerosol encounter probability (51.2% on the sitting sampling surface), but in the proposed design, this value was reduced to 3.5% (Nazari et al., 2022). Li et al. experimentally and numerically evaluated the impact of a low-cost solution, the box fan air purifier (BFAP), on direct and indirect aerosol transmission when placed in a small conference room with a high surface area-to-volume ratio. They found that the BFAP unit markedly reduced relative deposition beyond approximately 1 m from the breathing simulator, which can help to mitigate aerosol dispersion in enclosed spaces (Li et al., 2022).

While elevators are heavily used in high-rise buildings worldwide and are potential hotspots for COVID-19 transmission, few studies have investigated the airborne infection risk associated with different ventilation strategies inside an elevator cabin. Shao et al. conducted a study evaluating airborne transmission risk in three practical settings, including an elevator cabin. They introduced a local risk index and evaluated its spatial variation under different ventilation settings, noting that inappropriate ventilation design is a determining factor for indoor infection hotspots (Shao et al., 2021). However, this study evaluated only one type of ceiling ventilation with low aerosol removal efficiency (less than 20%). In another study, Dbouk et al. investigated the effects of mechanical ventilation systems and air purifiers on airborne virus transmission inside an elevator cabin and found that an air purifier only eliminates airborne transmission with a properly integrated ventilation system and air purifier design (Dbouk & Drikakis, 2021).

To date, no systematic study has investigated airborne transmission in an elevator cabin under different ventilation systems, and no guidance on the minimum ventilation rate associated with elevator capacity has been established. Thus, the present study was designed to systematically investigate the efficiency of different ventilation designs in reducing infection risk levels in elevators using CFD tools and to identify optimal design and minimum ventilation rate and time applied to minimize risks under different capacities. These findings will allow elevators to operate at normal capacity while minimizing airborne transmission risk. The following section presents the study methods, including modeling of the elevator cabin and spatial mesh, governing equations, and verification of results with multiple case studies. Section 3 represents the effects of various elevator HVAC designs on the aerosol probability infection risk considering the effects of capacity and air change rate. This study introduces novel criteria to show the effects of the streamline pattern of HVAC design on elevator performance. Additionally, the performance of each HVAC system on particle removal efficiency is calculated. The final section provides a summary of the findings and conclusions.

## 2. Methodology

### 2.1. Elevator cabin model and spatial mesh

A three-dimensional model of a 3,500-pound elevator with a maximum capacity of 21 persons was developed for the present study. The elevator cabin has overall dimensions of 2.03 (m) × 1.65 (m) × 2.44 (m) (L × W × H), as shown in detail in **Figure 1a**. The model represents a zone with nine standing locations divided into three rows. The minimum distance between each passenger is 0.39 m, and the maximum distance is 1.43 m, which is closer than the recommended 6 ft separation distance for COVID-19 mitigation (Niazi et al., 2021). All passengers are assumed to be represented similarly with the same dimensions. Each passenger is represented as a mannequin block with the dimensions of 0.28 (m) × 0.16 (m) × 1.8 (m) (L × W × H) and a total body surface area of 1.73 $m^2$ facing the elevator door, as shown in **Figure 1a**. The human body model follows the models used in numerical simulations of the dispersion of human exhaled droplets under different ventilation methods in a classroom (Farouk et al., 2021). The height of the mouth is 1.60 m above the floor, and the mouth opening is 3.14 $cm^2$, with a diameter of 10 mm. An asymptomatic person with COVID-19 (hereafter infector) is assumed among the passengers inside the elevator cabin.

A highly fine unstructured mesh is used with near-wall refinement, as shown in **Figures 1d and 1e**. The mesh consists of $5.6 \times 10^6$ mesh cells when nine passengers are standing inside the elevator cabin. With the capacity reduction, the refined mesh near the passenger is replaced by the bulk domain mesh, which is relatively coarser than that near body. The total mesh reduces to $0.95 \times 10^6$ mesh cells when only one passenger stands inside the elevator cabin. The generated mesh has a maximum mesh size of ~15 mm and a minimum cell size of ~1 mm. A much smaller mesh near the body surface, walls, ceiling, and floor is used to ensure the maximum $y^+ < 1$ during the simulation. To ensure the similarity of the computational mesh between different passengers, only one mesh is generated near the passenger, and different passengers have the same mesh refinement. In the current study, the inhalation of all passengers is ignored, and only the exhalation breathing activity of the infector is considered. The mesh near the mouth of the infector person is further refined to around 1 mm in mesh size to ensure the accuracy of breathing jet dynamics. To verify that the computations are sufficiently independent from the mesh, the results regarding the mesh size of $0.95 \times 10^6$ were compared with a much finer mesh with $1.2 \times 10^6$ mesh cells. The differences observed do not suggest the need for additional refinement of the base mesh. The pressure drop of the supply air of the elevator cabin passes through a porous media was calculated using the following equation:

$$\Delta p = \frac{1}{2} C_2 \rho . \Delta n . v^2 + \frac{\mu}{\alpha} . \Delta n . v \qquad (1)$$

where $\mu$ is the aerodynamic viscosity, which is set at $1.8 \times 10^{-5}$; $v$ is the air velocity; and $\Delta n$ represents the thickness of the supply panel. To obtain accurate results, a porous boundary condition was applied to the elevator supply air ducts. The values of $C_2$ and $\frac{1}{\alpha}$ were set at 150000 m$^{-1}$ and $1.5 \times 10^6$ m$^{-2}$, respectively (Nazari et al., 2022).

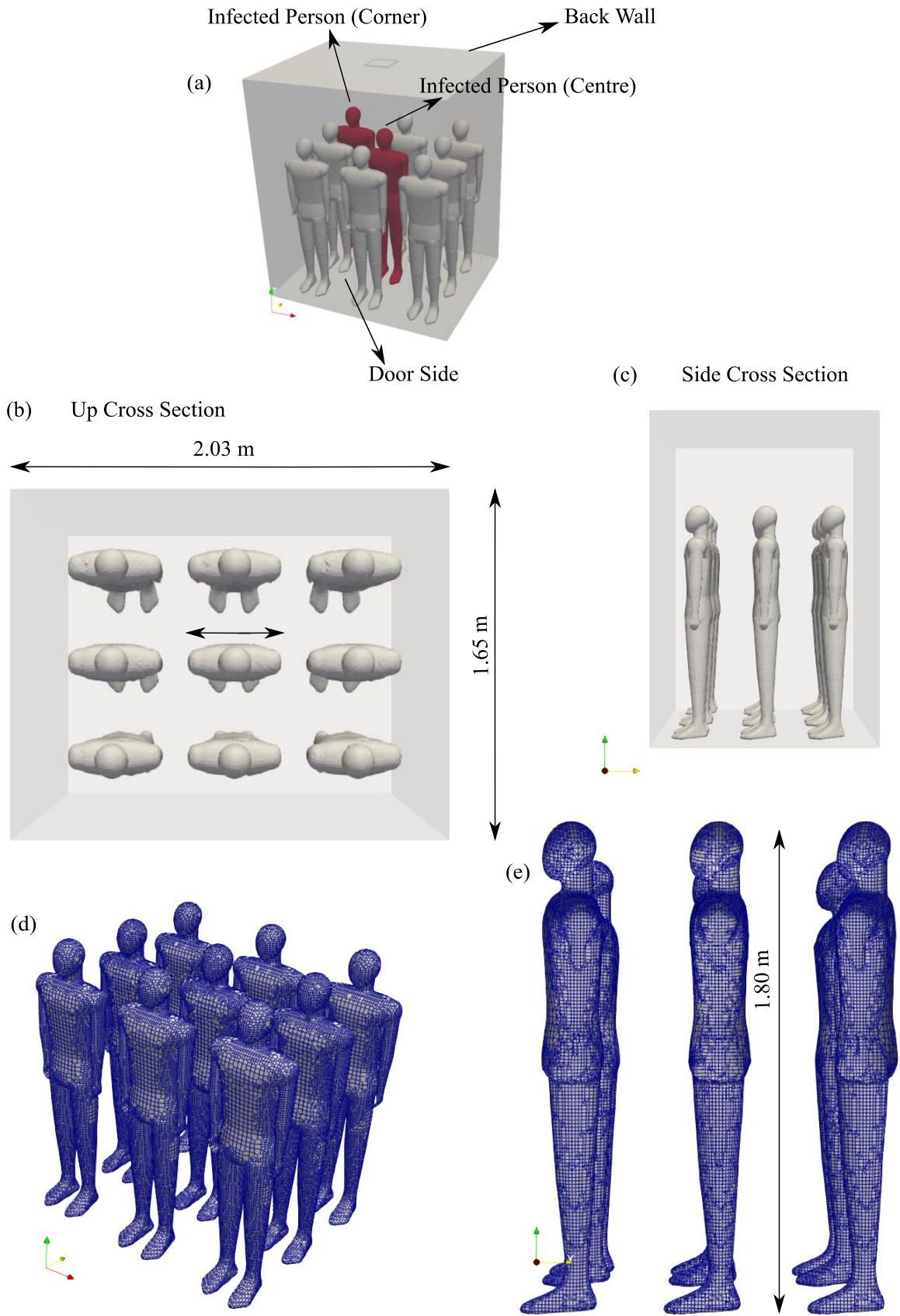

**Figure 1.** Illustration of (a) elevator cabin model, (b) up cross section, (c) side cross section, and (d and e) the computation mesh used in the CFD simulations.

## 2.2. Numerical simulations

The present study uses the open-source finite volume CFD code (*OpenFOAM*-5) to simulate airflow and aerosol transport in the Eulerian framework. Airflow is calculated by solving the incompressible Reynolds-averaged Navier–Stokes (RANS) equations, including mass conservation, momentum conservation, and energy conservation equations of the carrier phase, along with the specie equation using the Eulerian descriptions. The transmission of respiratory aerosols containing the virus was tracked by solving the translation equation of the concentration (Elghobashi, 1994; Bourouiba *et al.*, 2014). The following equations give the governing equations of the carrier phase (airflow). (Nazari et al., 2022; Nazari et al., 2021b; Wang and Hong, 2023):

$$div(\bar{u}_i) = 0 \tag{2}$$

$$\rho \frac{\partial}{\partial x_j}(\bar{u}_i \bar{u}_j) = -\frac{\partial \bar{p}}{\partial x_i} + g.x\nabla(\frac{\rho}{\rho_0}) + \frac{\partial}{\partial x_j}[\mu(\frac{\partial \bar{u}_i}{\partial x_j} + \frac{\partial \bar{u}_j}{\partial x_i})] + \frac{\partial}{\partial x_j}(-\rho \bar{u}_i' \bar{u}_j') \tag{3}$$

$$\rho c_p (\frac{\partial}{\partial x_j} \bar{u}_i \bar{T}) = \nabla.(k\nabla \bar{T}) + \frac{1}{2}\tau:(\nabla \bar{u}_i + \nabla \bar{u}_j^T) + \frac{\partial}{\partial x_j}(-\rho \bar{u}_i' \bar{T}) \tag{4}$$

$$\rho \frac{\partial}{\partial x_j}(\bar{u}_i \bar{C}) = \nabla.(D_{eff} \nabla \bar{C}) + \frac{\partial}{\partial x_j}(\rho \bar{u}_i' \bar{C}) \tag{5}$$

Where $\rho, \rho_0, p, u, C, T, k,$ and $D_{eff}$ are the density, reference density, dynamic pressure, velocity, aerosol concentration, temperature, thermal conductivity, and diffusivity of fluid, respectively. In equation (3), the *Boussinesq* approximation was used for the momentum equation. Also, in equation (5), $D_{eff}$ is $\frac{\nu_t}{Sc_t} + \frac{\nu}{Sc}$, where $Sc_t$ and $Sc$ are the turbulent and laminar Schmidt numbers, respectively, and $Sc_t = Sc = 1$. To ignore the Brownian motion effect on aerosol diffusion, it was assumed that the Schmidt number is equal to 1. This means that aerosols diffuse at the same rate as the momentum for 1–10 $\mu m$ particles in the breathing model. Considering the dilute particle concentration, the particle-induced mass, momentum, and energy transfer were ignored. The second-order upwind scheme was used to handle the convective terms, and the Gauss-linear second-order approach was used for the diffusion terms. The non-orthogonality correction of the computational cells was applied for the gradients. The Pressure-Implicit with Splitting of Operator (PISO) algorithm was applied to couple the pressure and the velocity. The maximum residuals for the convergence of pressure and velocity were $10^{-8}$ and $10^{-12}$, respectively. The standard k-ε turbulent model was used for all simulations in this study. This model has been shown to be reliable for modeling airflow in enclosed spaces, and its computational cost is low, as suggested by Nazari et al. (2022) and Gao et al. (2017). The simulation model was based on the following assumptions: (1) The size and physical properties of the respiratory aerosol particles remain constant throughout the simulation; (2) The generation of respiratory particles is not considered; (3) The effects of spike-like structures located at the viral surface on viral transmission are negligible; (4) The elevator cabin acceleration is assumed to be constant.

**2.3. Study design**

**Figure 2** summarizes the simulation cases presented in the current study, including (1) mixing ventilation (MV-1, MV-2, MV-3, MV-4, MV-5, MV-6, and MV-7), (2) displacement ventilation (DV-1), and stratum ventilation (SV-1 and SV-2). As a commonly used mechanical ventilation mode, mixing ventilation (MV) dilutes virus concentration by thoroughly mixing with the air. DV can provide an advantage in reducing aerosol concentration based on the principle of displacing contaminated indoor air with fresh air from outside. For base cases, the elevator cabin is equipped with a mixing ventilation system (MV-1) selected from commercially available ventilation systems inside the elevator cabin, as shown in **Figure 2a** (Du & Chen, 2022). The inlet of the ventilation system is located at the ceiling, while several distributed outlets are on sidewalls near the floor. The inlet and outlet dimensions of MV-1 are 0.4×0.4 m$^2$ and 0.15×0.1 m$^2$, respectively. An outlet pressure boundary condition is applied at the MV-1 outlet, while a constant velocity boundary condition is used for the MV-1 inlet.

The velocity of the MV-1 was set as 0.10 m/s, corresponding to 34 CFM and 7 ACH for the simulated elevator cabin size, which is used to add the same amount of outside clean air with the recycled polluted air. No-slip boundary condition was applied to solid walls. The temperature inside the elevator cabin was set as 20 °C, and an adiabatic boundary condition was used for the elevator cabin sidewalls, floor, and ceiling to prevent heat transfer from the outside. The temperatures of the mannequin's surface and respiratory flow were set to 31 °C and 33 °C, respectively (Nazari *et al.*, 2022). In the present study, the heating or cooling effects of the ventilation system were ignored. Therefore, the ventilation system temperature at the inlet is the same as the room temperature. An infected person stands at two locations in the center or the corner. The passengers were considered the only heat sources driving a thermal buoyancy flow in the elevator cabin. The breathing activities were modeled by applying a time-varying injection profile at the person's mouth to mimic human breathing. The respiratory flow rate was set to 0.00033 m$^3$s$^{-1}$ based on experimental data for all the simulation cases (Nazari *et* *al.,* 2022). The particles were injected at 44 particles per breathing cycle. The simulations were conducted over a six-minute duration for particle injection, representing an infected person and elevator HVAC operation taking the same time. It was suggested that the elevator has a constant gravitational acceleration of 9.81 m/s$^2$. In other words, the movement direction of the elevator cabin effects is neglected.

To study the ventilation flow direction effects, two case studies with an outward flow design in the inlet at the ceiling (opposite to the previous cases at the base) were designed. To investigate the performance of different ventilation systems, another nine ventilation systems covering mixing ventilation (MV-2, MV-3, MV-4, MV-5, MV-6 and MV-7), displacement ventilation (DV-1), and stratum ventilation (SV-1 and SV-2) were adopted (**Figure 2**). Given two infection sources (passengers 1 and 5) and nine ventilation systems, sixteen cases were simulated in this study. For the enhanced ventilation rate effects, the velocity of the ventilation system was increased to achieve an increase in effective air changes from 2 to 12 ACH. To further examine the capacity effects, the cases were simulated with different capacities (two, five, and nine passengers) inside the elevator cabin. Besides, the particle persistence inside the elevator cabin was also evaluated. In this case, the persistence of contaminated particles was analyzed while no passenger was inside the elevator cabin, and the ventilation system was in operation.

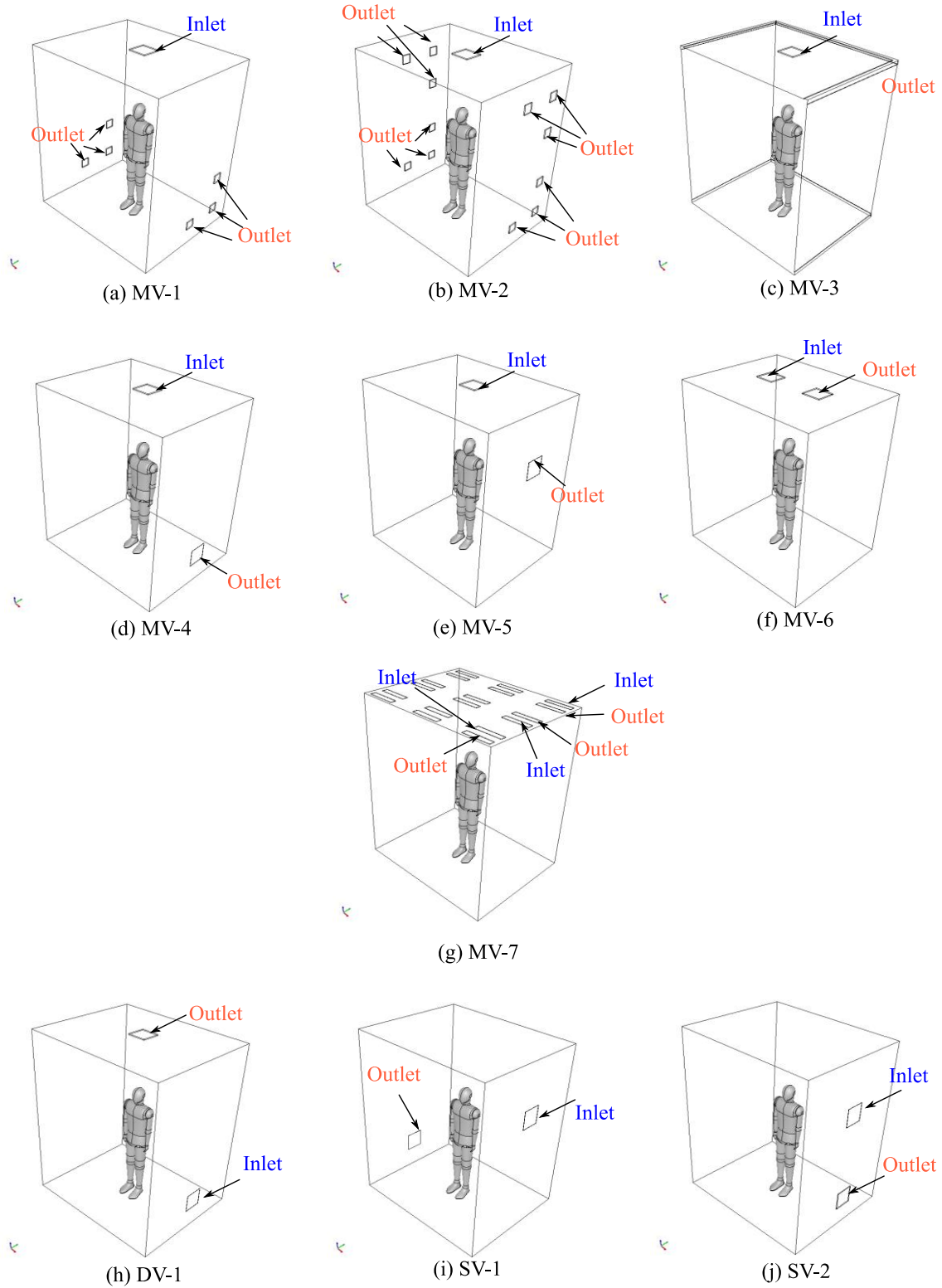

**Figure 2.** Schematic diagrams of different ventilation systems inside the elevator cabin showing the location of supply diffusers (inlet, blue), exhaust grilles (outlet, red), and one breathing thermal mannequin standing in location 5 (center) in Figure 1. Abbreviations: MV, mixing ventilation; DV, displacement ventilation; SV, stratum ventilation.

## 2.4. Simulation validation

In this investigation, *Salome* software (version 6) and *SnappyHexMesh* utility were applied to generate and discrete the mannequins' bodies, respectively. Due to the complexity of the geometrical model, this study used the unstructured grid adapted to various geometric structures. The results of the mesh-independent test are shown in **Figure 3.** The suitable grid for the one mannequin case was 0.95 million cells. This value for five and nine mannequins was 1.7 million and 3.9 million cells, respectively. **Figure 3d** shows the three configurations of meshes for one mannequin case, and **Figure 3e** indicates that the 0.95 million mesh is suitable for the present work. The variation of $I_{risk}$ over the elevator length is indicated in **Figure 3e.** The difference between 0.95 million and 1.2 million lines at different locations is nearly negligible, as shown in the figure. To validate the applied numerical method to calculate aerosol transmission, the CFD simulations were compared with Zhang *et al.*'s study for an enclosed room with displacement ventilation (Zhang et al., 2006). **Figures 3b and 3c** compare the two studies regarding velocity and temperature distributions, respectively. *Zhang et al.* (2006) used experimental measurements and numerical simulations to observe the particle transport and distribution in ventilated rooms. Seven sampling points were defined to compare the numerical and experimental results. They applied the human simulator to determine the interaction between the complex flow of the HVAC system and the source of pollutants. The calculated points were positioned in the middle of the enclosed room and were plotted in the *z*-direction. In this study, the velocity and temperature in the enclosed room were calculated at three vertical measurement points and compared with the experimental work of Zhang *et al*. (2006). Then, the maximum differences in the vertical velocities and temperatures were calculated and compared (P1, P4, and P7 points of **Figure 3a**). The maximum relative deviations of the velocity and temperature for the two cases in **Figures 3b and 3c** were approximately 1.9% and 0.3%, respectively.

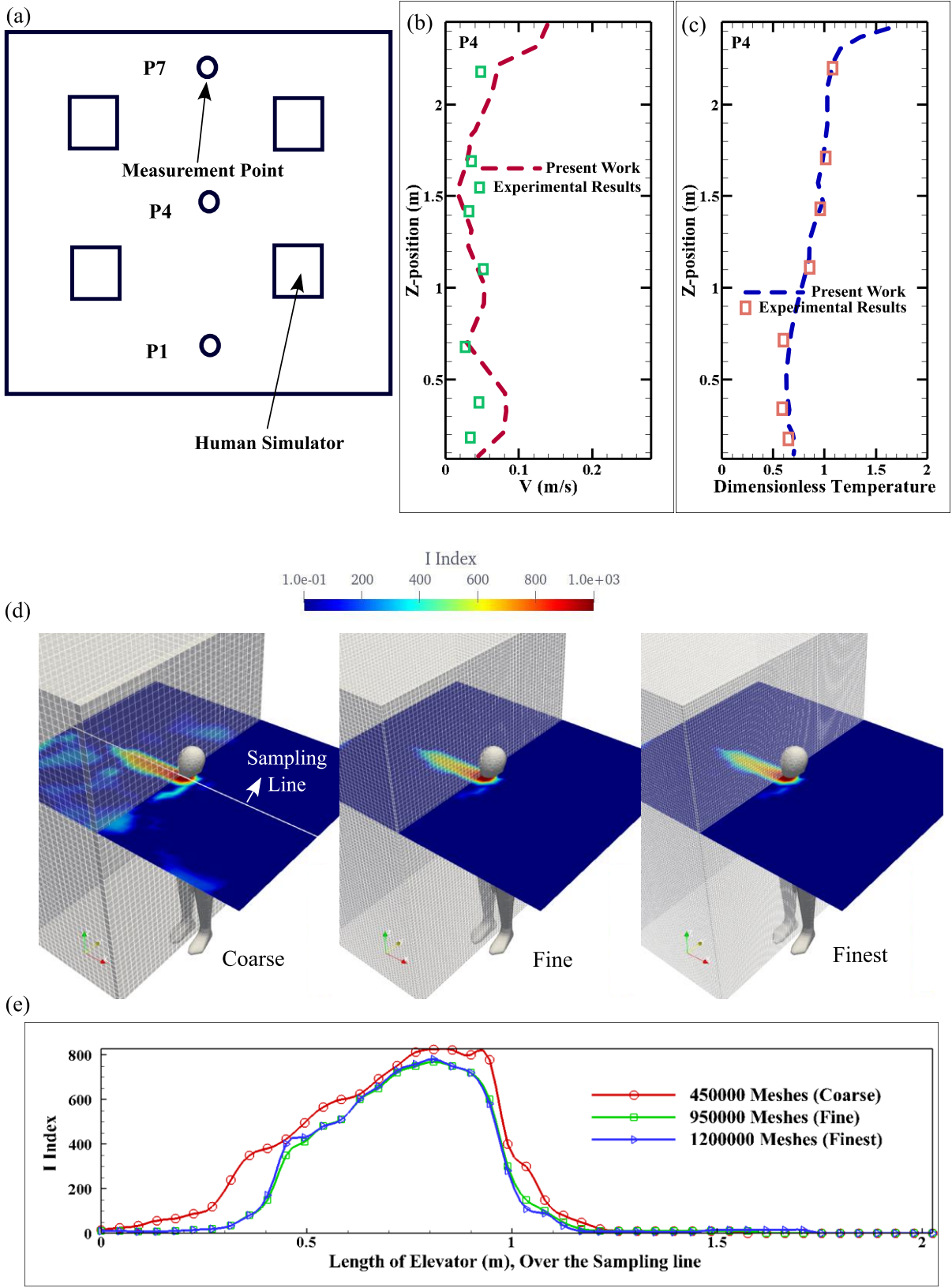

**Figure 3.** (a): Schematic views of human simulators and sampling points per Zhang et al. (2006) study. (b) and (c): Comparison of the velocity and temperature over the P4 with Zhang *et al.*'s experimental results (Zhang et al., 2006). (d) and (e): $I_{risk}$ contour and variations in front of mouth versus position along the elevator length (sampling line) obtained using a coarse mesh of 0.45 million cells, the fine mesh of 0.95 million cells, and the finest mesh of 1.2 million cells, respectively.

## 2.5. Criterions of elevator HVAC designs evaluation

### 2.5.1. Probability of infection risk

To quantify the relative risk of virus-containing particles at a given location, the aerosol probability index (P) was applied. Based on the typical temperature and ventilation velocity in an elevator, the Wells–Riley equation, $P = 1 - e^{\frac{-qt}{I_{risk}}}$ has been used to calculate the probability of infection risk. $I_{risk}$ (the risk index) is characterized by the risk of virus-containing particles at a sampling surface. This metric quantifies the number of particles passing through a specific location during the simulation period. It is calculated by summing the total number of particles passing through each point on the sampling surface throughout the entire simulation duration.

$$I_{risk} = \sum Q_i \tag{6}$$

where $Q_i$ is defined as:

$$Q_i = \begin{cases} 1, & \text{the first time the ith particle appears at sampling surface} \\ 0, & \text{otherwise} \end{cases}$$

The sampling surfaces cross the centers of the passenger's breathing zone hemispheres (**Figures 4a and 4c**). "We defined the hemispherical breathing zone with a radius of 300 mm, which encompasses the area around the nose and mouth from where the majority of air is drawn into the lungs. The quantum generation rate for an infected person (q) is expressed in units of quanta per second (quanta. s$^{-1}$). A higher value of $I_{risk}$ in a particular zone indicates a higher risk of transmission. The number of ejected particles is defined as

$$N = \sum_0^t C\dot{V} \tag{7}$$

Where $t$ and $\dot{V}$ are the exposure time and human breathing rate set to 240 s and 0.00033 m$^3$.s$^{-1}$, respectively (Nazari *et al*., 2022; Zhang *et al*., 2021b). In this study, it was assumed a continuous flow at the breathing sources inside the elevator cabin (Nazari *et al*., 2022; Zhang *et al*., 2021b). Besides, it was assumed that the air at the individuals' mouths is constantly outward at a rate of 0.2 times. s$^{-1}$ (respiratory period of ~5 s and 12 breathing per minute). The simulation time is sufficiently long to neglect periodic effects as the focus is on airborne transmission far from the emission source. A continuous breathing model is suitable for analyzing virus movement inside the elevator cabin over the exposure time. While the pulsatile nature of breathing was ignored in this study, a steady flow velocity equivalent to the average velocity of the entire respiratory period was applied. This period is essential for exhalation-based viral load and simulating aerosol dispersion in a confined space. To obtain a steady equivalent result, the oscillatory velocity flow index was considered, which averages the flow velocity throughout the simulation from the start to the end of the exhalation phase. An analogous form of the oscillatory index (OI) is expressed in the equation below (Piemjaiswang et al., 2019):

$$OI\{\varphi\} = 1 - \frac{\left\|\frac{1}{T}\int_T \varphi \, dt\right\|}{\frac{1}{T}\int_T \|\varphi\| \, dt} \tag{8}$$

Where $\varphi$ is the generic flow field variable and $T$ is the assessment time. By weighting the oscillatory index with flow velocity, the oscillatory velocity flow index (OVFI) is obtained, as shown in the equation below (Piemjaiswang *et al.,* 2019):

$$OI\{u\} = OVFI = 1 - \frac{\|\bar{u}\|}{\|u\|} \tag{9}$$

### 2.5.2. Suppression effect (SE)

Another criterion used in the present work is the suppression effect (SE). This study introduced the HVAC suppression effect to represent the persistence of particles below the breathing zone during airborne transmission.

$$SE = \frac{\sum_0^{t_{ES}} Q_{SE}}{Q_{all}} \tag{10}$$

In this study, the portion of aerosols that enter under the breathing zone at least once and then either rapidly escape from the elevator by exhausted ducts or remain in that zone is defined as the $Q_{SE}$. HVAC designs of MV-1, MV-2, MV-3, MV-4, SV-1, and SV-2 have exhaust ducts under the breathing zone, and the recirculated aerosol increases the risk index over the sampling surface. To better understand the effect of HVAC duct configuration on aerosol behavior, the present work introduces the concept of the SE zone, which is shown in Figure 4b, while Figures 4a and 4c depict the sampling surface and breathing zone, respectively. The SE zone is calculated using the *Swak4Foam* utility in *OpenFOAM* and demonstrates the effect of the streamlines of HVAC design on respiratory behavior. This concept was recently proposed by Nazari et al. for a novel HVAC system in urban subways (Nazari et al., 2022), and to the best of the authors' knowledge, its suitability for long and large spaces such as elevators has not been studied. In elevators, the no-slip boundary condition of the walls in the small space can change the flow direction of the aerosol carrier fluid and cause the aerosol to disperse inside the elevator. Thus, the present work investigates the effect of various elevator HVAC designs on aerosol dispersion.

**2.5.3. Particle removal efficiency ($\varepsilon_p$)**

Another important criterion for evaluating HVAC performance in an elevator is particle removal efficiency ($\varepsilon_p$). This parameter is defined as the ratio of particles that are escaped from ventilation outlets to the total particles injected inside the elevator under different ventilation rates. This particle removal efficiency has been commonly used to evaluate ventilation performance for mitigating airborne transmission.

$$\varepsilon_p = \frac{\sum_0^t Q_{Exit}}{Q_{all}} \qquad (11)$$

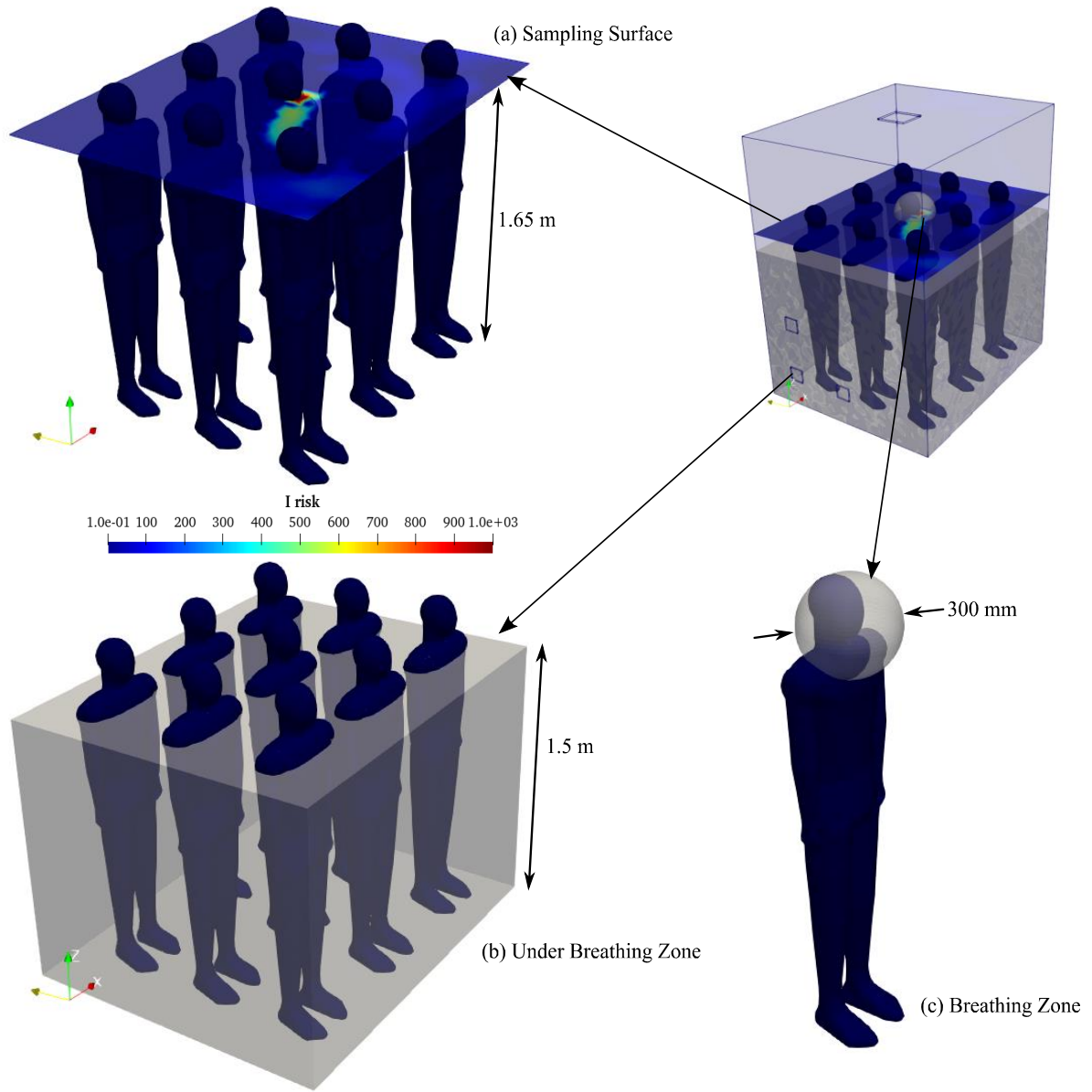

**Figure 4.** Schematic view of breathing zone, under breathing zone, and sampling surface.

## 3. Results and Discussion

In this section, the concentration contours for multiple scenarios at exposure time are presented. It should be noted that irregular stopping of the elevator cabin to load and unload passengers is a significant feature of elevator transport, which further increases the risk of viral transmission. The exposure time depends on the height of the building. The present work focuses on the aerosol dynamics inside the elevator cabin by illustrating the effects of different ventilation designs. Based on the accumulation of respiratory particles, the results indicate that the MV-7 design can effectively suppress aerosol clouds at both locations of the infected individual due to the regular streamline under each duct, which prevents aerosol dispersion.

Additionally, the DV-1 HVAC design showed higher particle removal efficiency and lower particle accumulation on the sampling surface compared to SV-1 and SV-2 designs.

In the second step of the study, the impact of elevator capacity on the infection risk was investigated. The findings revealed that as the elevator capacity increased, the infection risk probabilities for both DV-1 and MV-7 HVAC designs also increased. The MV-7 HVAC system was observed to be more effective in quickly removing aerosols through a safe pathway, owing to the separated ventilation over each individual. However, as the number of individuals inside the elevator increased, the pathway for particles to exit decreased, which led to an accumulation of particles on the sampling surface. This accumulation can lead to an increased risk of infection, especially in poorly ventilated elevators. In the third step, the effect of elevator ACH on breathing aerosols was investigated. Three ACHs, namely 2, 7, and 12 $hr^{-1}$, were chosen for this part of the study. The results showed that increasing the ACH from 2 to 12 reduced the aerosol infection risk for both center- and corner-located infector cases. Poorly ventilated elevators, with low ACH values, create weak shear flow unable to change the aerosol cloud's direction, leading to a higher risk of virus transmission towards exhaust ducts. In poorly ventilated scenarios, a zone with high particle accumulation levels was formed, corresponding to locations with higher virus transmission risks. **Figure 5** summarizes the three main steps of the study.

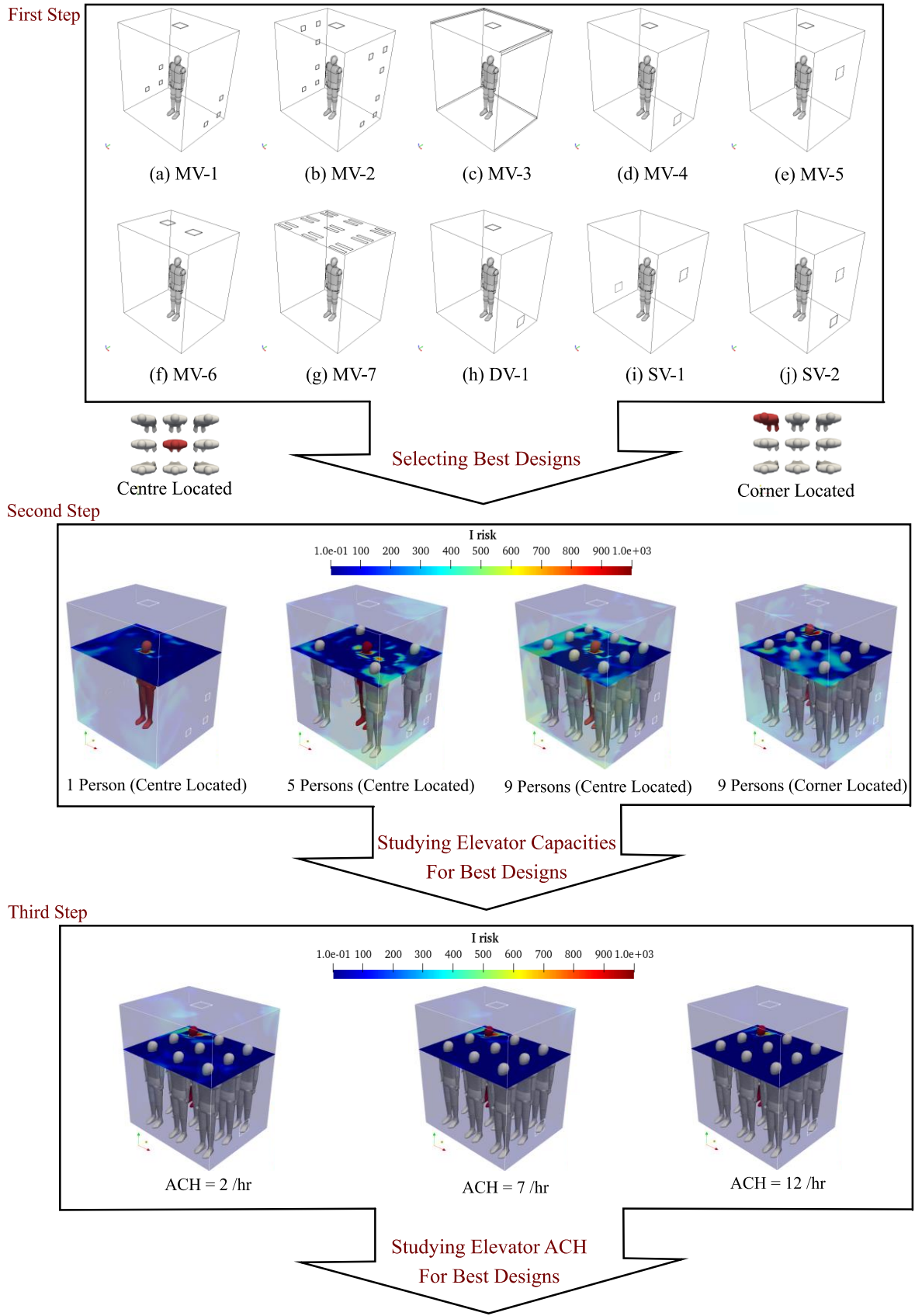

**Figure 5.** Schematic view of three main steps of the present work. The first step demonstrates selecting an appropriate HVAC design based on respiratory particle distribution on the sampling surface. In the second step, the effect of elevator capacity on the respiratory particle accumulation over the sampling surface was evaluated. The third step shows the effects of ACH on particle movement.

### 3.1. Aerosol infection risks under different ventilation designs

**Table 1** provides a summary of the calculated criteria for various HVAC designs in the present study, presented for both center- and corner-located infectors. The results show that the MV-7 design has the highest particle removal efficiency, with maximum values of 75.90% and 79.40% for center- and corner-located infectors, respectively. For MV-6 and MV-3, the particle removal efficiency for center-located infectors is 62.40% and 13.20%, respectively, while the corresponding values for corner-located infectors are 43.01% and 16.84%, respectively. Among the mixing ventilation designs, MV-3 had the lowest particle removal efficiency due to its corner outlet ducts and opposing inlet flow direction to gravity, leading to cross-contact between upward outlets that reduces particle removal efficiency. For DV-1, SV-1, and SV-2 HVAC designs, the calculated particle removal efficiency for center-located infectors were 74.79%, 7.93%, and 4.99%, respectively, while the corresponding values for corner-located infectors were 42.77%, 4.90%, and 3.97%, respectively. Among the three designs, DV-1 had the highest particle removal efficiency due to the shorter pathway for particle removal, reducing suppression effects.

The stratum ventilation designs (SV-1 and SV-2) showed the highest suppression effect among all designs for both center and corner-located infectors, meaning that they were able to remove the maximal number of particles outside the breathing zone, i.e., under the breathing zone.

**Table 1.** The summary of calculated criteria in the present work.

| Investigation | Ventilation type | Source | ACH | Suppression Effect (%) | Particle removal efficiency (%) | Probability of infection risk (%) over the sampling surface |
|---|---|---|---|---|---|---|
| Base case | MV-1 | Center | 7 | 53.20 % | 12.31% | 16.20 % |
|  | MV-1 | Corner | 7 | 58.30 % | 10.85% | 11.80 % |
| Effect of air distribution | MV-2 | Center | 7 | 22.20 % | 16.62% | 19.50 % |
|  | MV-3 | Center | 7 | 28.30 % | 13.20% | 10.00 % |
|  | MV-4 | Center | 7 | 62.10 % | 16.93% | 16.80 % |
|  | MV-5 | Center | 7 | 26.70 % | 20.41% | 30.20 % |
|  | MV-6 | Center | 7 | 10.40 % | 62.40% | 07.00 % |
|  | MV-7 | Center | 7 | 05.10 % | 75.90 % | 06.10 % |
|  | DV-1 | Center | 7 | 13.20 % | 74.79% | 18.80 % |
|  | SV-1 | Center | 7 | 65.40 % | 07.93% | 13.00 % |
|  | SV-2 | Center | 7 | 73.40 % | 04.99% | 14.80 % |

| | | | | | |
|---|---|---|---|---|---|
| MV-2 | Corner | 7 | 30.40 % | 15.61% | 35.20 % |
| MV-3 | Corner | 7 | 35.10 % | 16.84% | 16.00 % |
| MV-4 | Corner | 7 | 58.00 % | 10.17% | 20.20 % |
| MV-5 | Corner | 7 | 22.70 % | 19.14% | 31.70 % |
| MV-6 | Corner | 7 | 08.20 % | 43.01% | 10.50 % |
| MV-7 | Corner | 7 | 05.30 % | 79.40 % | 07.30 % |
| DV-1 | Corner | 7 | 18.40 % | 42.77% | 09.00 % |
| SV-1 | Corner | 7 | 67.70 % | 04.90% | 07.10% |
| SV-2 | Corner | 7 | 62.40 % | 03.97% | 19.05 % |

Abbreviations: MV, mixing ventilation; DV, displacement ventilation; SV, stratum ventilation.

In **Figure 6**, the infection risk distribution under different mixing ventilation designs for a center-located infection source is presented. The results show that the distribution of infection risk for MV-6 and MV-7 is lower than for other ventilation designs. However, the highest infection probability among the mixing ventilation systems belongs to MV-5, with a value of 30.2%. This could be due to the exhaust duct position increasing aerosol movement over the sampling surface, which raises the infection risk for each person inside the MV-5 design. To further investigate the physical mechanisms involved in the SE and infection risk probability for the center-located infector responsible for the observations in **Figure 6**, the corresponding streamlines are illustrated in **Figure 7**. The positions of ducts and streamlines directly influence the suppression effect values. As shown in **Figure 7**, the streamlines are characterized by the positions of ducts, and flows generated by the HVAC system can affect infection risk. The calculated probability of infection risk for MV-7 for the center- and corner-located infectors was 06.10% and 07.30%, respectively. However, the relevant SE for the mentioned locations was 05.10% and 05.30%, respectively. This indicates that the MV-7 design mainly exits the created particle from the top part of the elevator cabin, and the flow pattern does not allow for aerosol to disperse under the breathing zone. Thus, this HVAC design is not recommended for situations involving loading and unloading of elevator passengers. The MV-6 and DV-1 designs have similar scenarios with minimal suppression effects. However, these HVAC designs are suitable for traveling elevators due to their lower risk of infection and efficient particle removal. In **Figure 8**, the infection risk distribution and flow streamline of DV-1, SV-1, and SV-2 designs are shown. The infection risk for SV-1 and SV-2 has the highest distribution over the sampling surface.

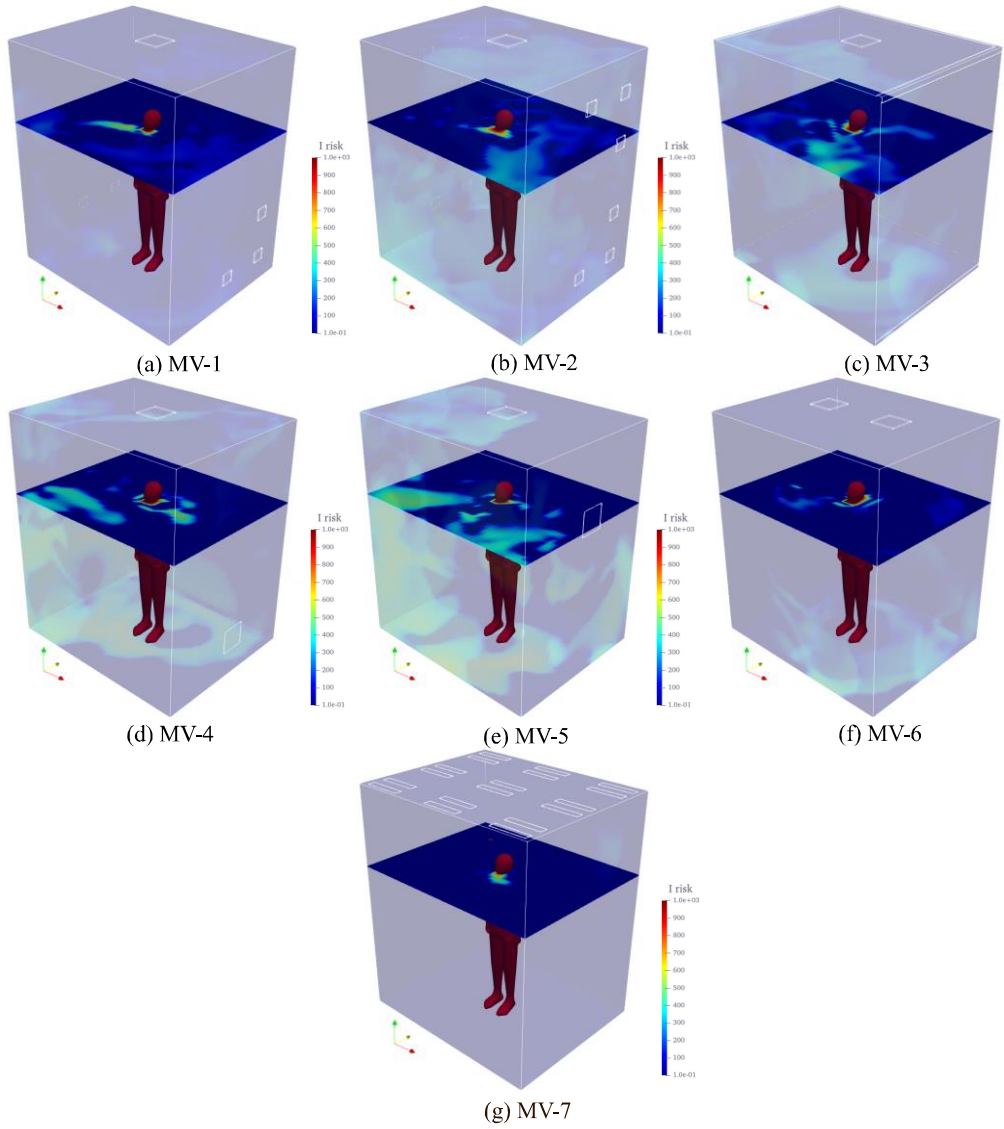

**Figure 6.** Influence of mixing ventilation designs on particle dispersion for center-located infector; ACH of 7 hr$^{-1}$

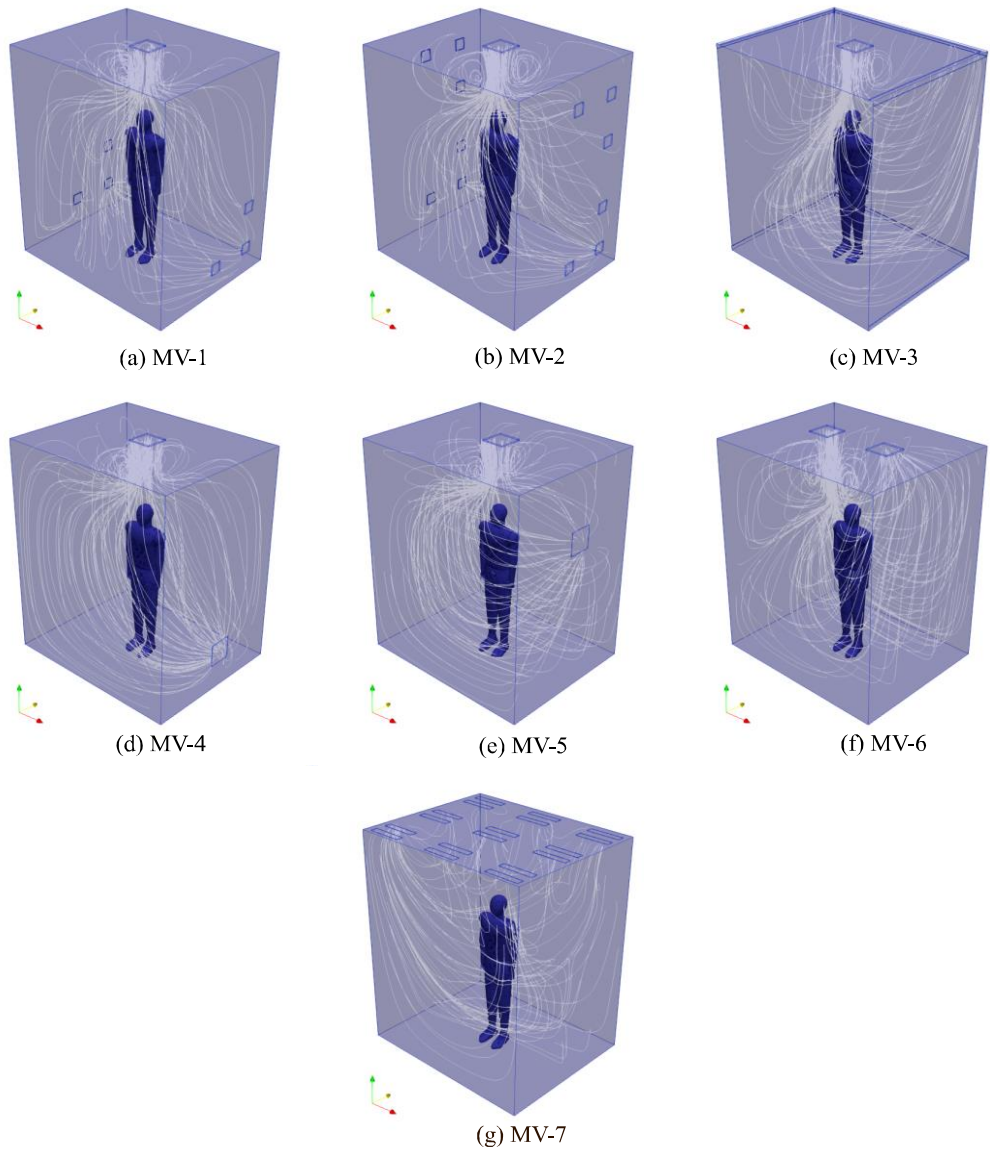

**Figure 7.** Influence of mixing ventilation designs on flow fields for center-located infector; ACH of 7 hr$^{-1}$

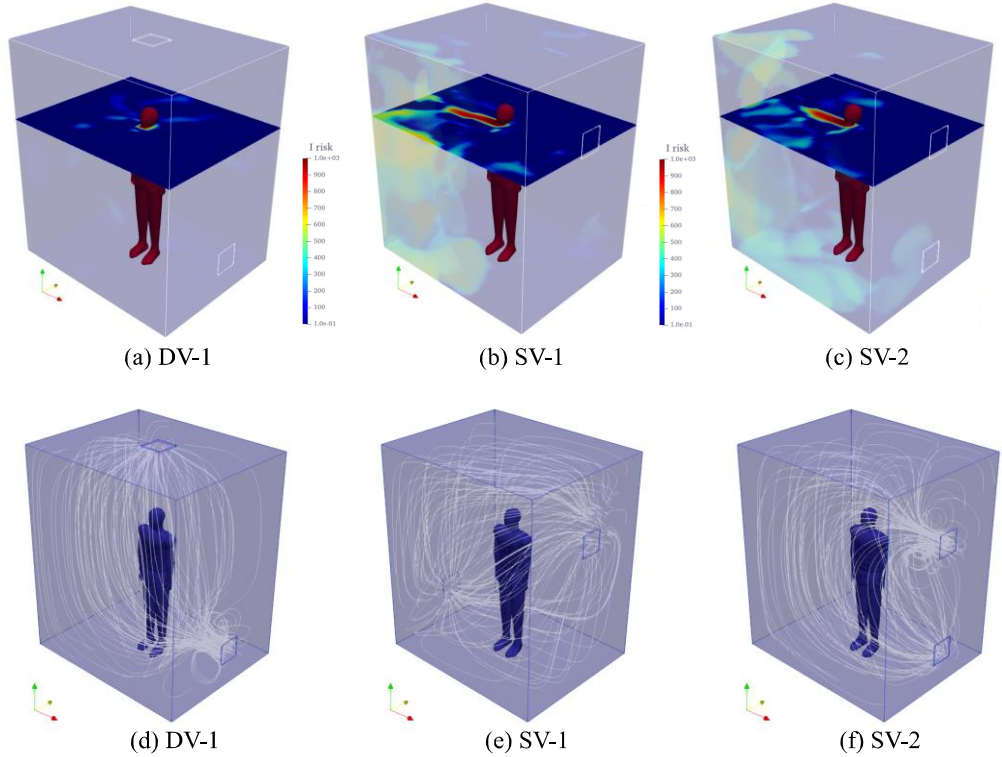

**Figure 8.** Influence of displacement and stratum ventilation designs on particle dispersion (top row) and flow field (bottom row) for center-located infector. ACH of 7 hr$^{-1}$

**Figure 9** presents the infection risk distribution under various mixing ventilation designs for a corner-located infector. Compared to the center-located infector conditions shown in **Figure 6**, the MV-7 design exhibits a higher infection risk distribution at the corner-located infector, resulting in a higher value of $\varepsilon_p$. Similarly, for all MV-2, MV-3, MV-4, MV-5, and MV-6 designs, the infection risk at the corner-located infector is higher than at the center-located infector. **Figure 10** displays the flow streamline under different mixing ventilation designs for a corner-located infector. **Figure 11** presents the infection risk distribution and flow streamline for DV-1, SV-1, and SV-2 designs.

The HVAC flow of MV-1 generates jittery streamlines due to the presence of multiple exhaust ducts distributed along the side walls. These jittery patterns intensify in the HVAC design of MV-2, where streamlines are divided into two directions. In these mentioned HVAC systems, the probability of infection risk escalates due to rapid changes in flow direction and jittery patterns. The study's results reveal that the HVAC flow in MV-3 is similar to MV-2 but with fewer jittery patterns. This HVAC configuration produces a smooth flow pattern near exhaust ducts and exhibits satisfactory particle removal efficiency, owing to the interaction of upward flow and effective HVAC design. In contrast, MV-4 and MV-5 possess flow patterns unsuitable for elevator HVAC design. MV-5 demonstrates an increased risk of infection due to aerosol exhaust near the mannequin breathing zone, while in MV-4, the suppression effect of HVAC and wall interactions create an undesirable flow pattern. The study concludes that smooth streamlines without vortexes in the breathing zone are essential for elevator HVAC design, which is unattainable in any of the tested models (MV-1 to MV-5), as all HVAC designs exhibit similar conditions in terms of inlet mass flow rate and exhausted duct area. Although the MV-6 design can control aerosols with rapid capturing, the large vortex size causes some aerosols to circulate near the side walls, heightening the probability of infection. In contrast, the HVAC design of MV-7 resolves this issue by implementing smooth inlet and outlet ducts above each individual. This arrangement generates smaller vortices above each person, yielding a smooth fluid flow pattern with a lower probability of infection and enhanced particle removal efficiency.

The fluid flow of DV-1 can transports aerosols upward in a smooth and non-turbulent manner, but its particle removal efficiency is superior to SV-1 and SV-2 due to its effective aerosol capturing. SV-1 sweeps aerosols from one side to the other without generating a vortex; however, since the transfer path is at the breathing zone, more than half of the aerosols exhibit a higher probability of infection across the sampling surface. For SV-2, fluid flow interacts with the front wall before returning to the exhaust duct on the same wall as the inlet, increasing the probability of infection and decreasing particle removal efficiency. Our research suggests that enhancing suppression effects may be suitable for larger spaces, such as urban subways. Nonetheless, in smaller enclosed spaces, this effect may introduce jittery patterns.

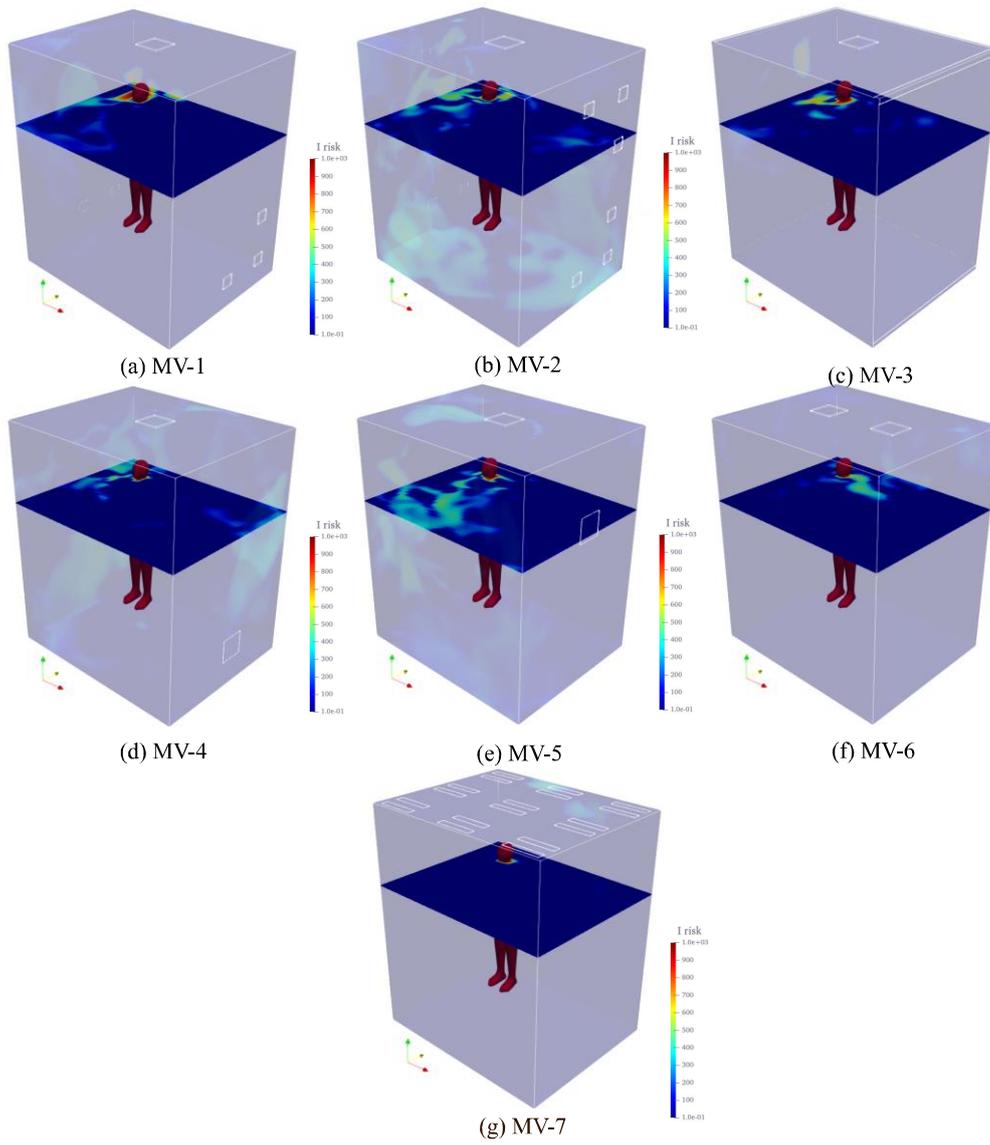

**Figure 9.** Influence of mixing ventilation designs on particle dispersion for corner-located infector; ACH of 7 hr$^{-1}$

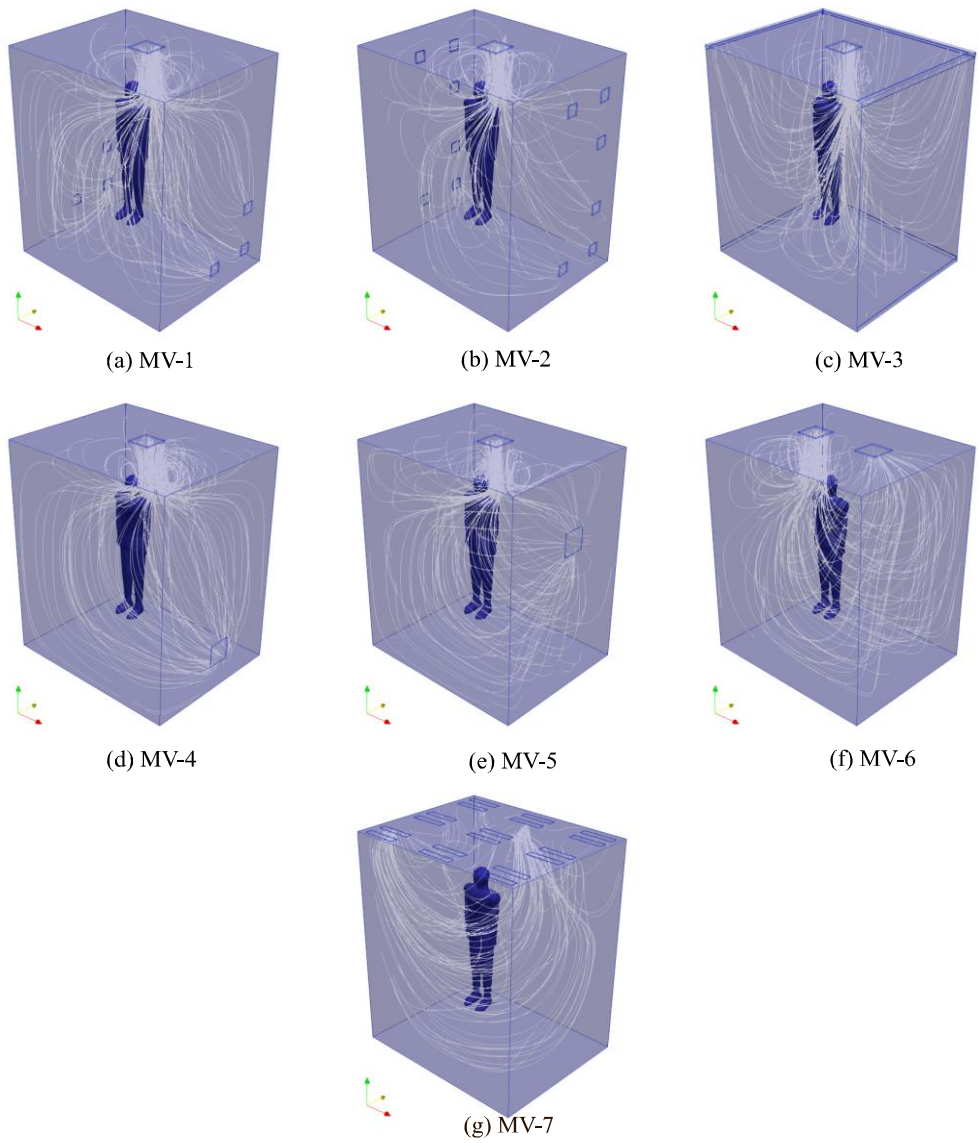

**Figure 10.** Influence of mixing ventilation designs on flow fields for corner-located infector; ACH of 7 hr$^{-1}$

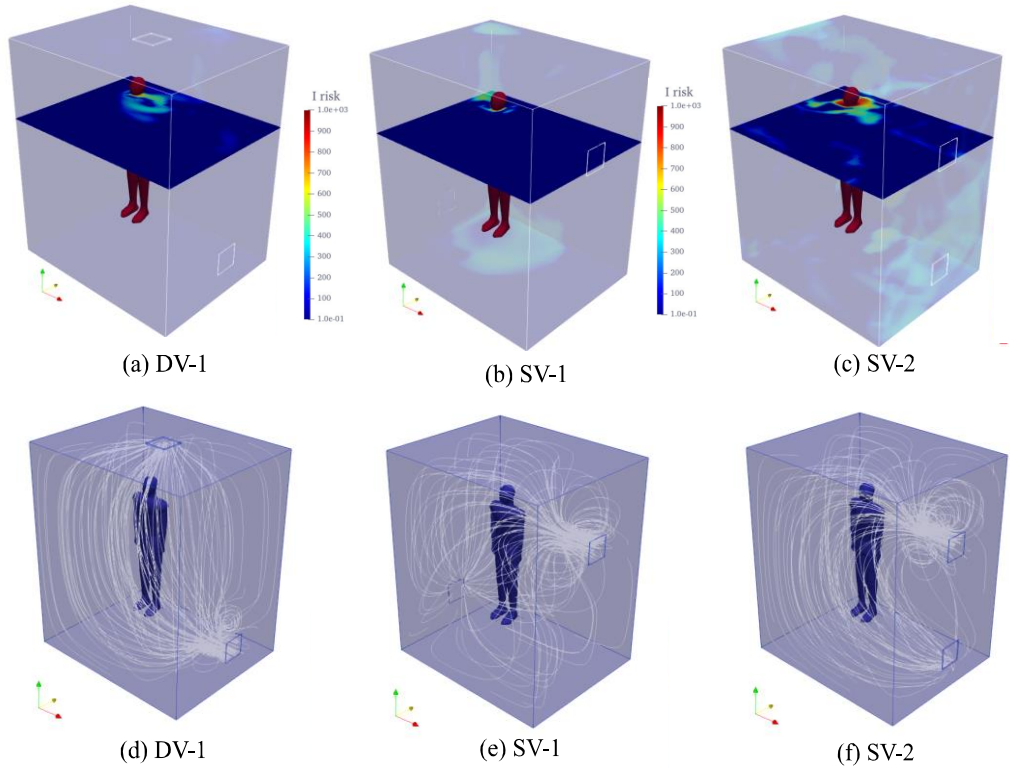

| (a) DV-1 | (b) SV-1 | (c) SV-2 |
| (d) DV-1 | (e) SV-1 | (f) SV-2 |

**Figure 11.** Influence of displacement and stratum ventilation designs on particle dispersion (top row) and flow field (bottom row) for center-located infector; ACH of 7 hr$^{-1}$

### 3.2. Influence of elevator capacities under different ventilation designs

The exposure time in elevator ventilation systems depends on the number of infected persons inside the elevator and the building height. The number of persons in the cabin is a critical factor in particle movement. To study the viral concentration inside the elevator, different elevator capacities (with two, five, and nine persons) were simulated, with the thermally active surface of any individual inside the elevator considered. With increased passenger numbers, particle removal becomes slower due to increased aerosol circulation inside the elevator resulting from interactions with individual passengers. The type of ventilation duct configuration and direction of particle movement are crucial to suppressing the pollutant. Appropriate HVAC designs (DV-1 and MV-7) were selected to evaluate the capacities. **Figures 12 and 13** show the infection risk distribution of MV-7 and DV-1 designs for two, five, and nine capacities. In high-capacity density scenarios, particles tend to remain localized near the infector, increasing the infection risk for healthy individuals nearby. In displacement ventilation designs, the outlet duct and infector position are crucial during elevator ascent to keep respiratory aerosols away from healthy breathing zones. However, the airflow pattern around the outlet duct may still disperse the aerosols in the breathing zone.

Under mixing ventilation, particles are dispersed throughout the domain, with the highest concentration near the infector. As the capacity and ventilation rate increase, particle removal efficiency decreases under both mixing- and displacement ventilation. High capacity with high ventilation rates show higher particle removal efficiency compared to low ventilation rates and low capacity. **Table 2** presents the calculated criteria for elevator capacity, indicating that increasing the capacity leads to an increase in the suppression effect of DV-1 and MV-7 HVAC designs. This slowed airflow can transfer a portion of respiratory aerosols to under the breathing zone, reducing the particle removal efficiency due to top located exhaust ducts in MV-7 and DV-1 designs. The maximum particle removal efficiency is 79.40% for one corner-located passenger for the MV-7 model, while for DV-1 and center-located passengers, it is 74.79%. However, the infection risk over the sampling surface increases with an increasing elevator capacity due to the entry of a portion of aerosols under the breathing zone.

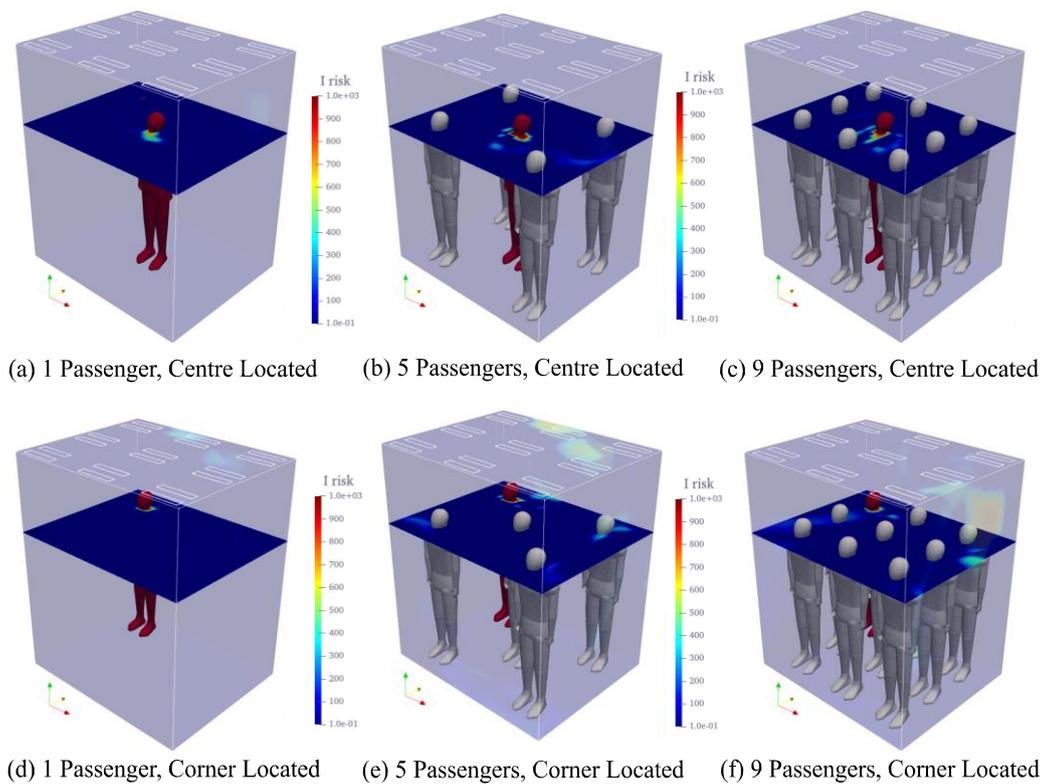

(a) 1 Passenger, Centre Located  (b) 5 Passengers, Centre Located  (c) 9 Passengers, Centre Located

(d) 1 Passenger, Corner Located  (e) 5 Passengers, Corner Located  (f) 9 Passengers, Corner Located

**Figure 12**. Influence of capacities on particle dispersion for MV-7.

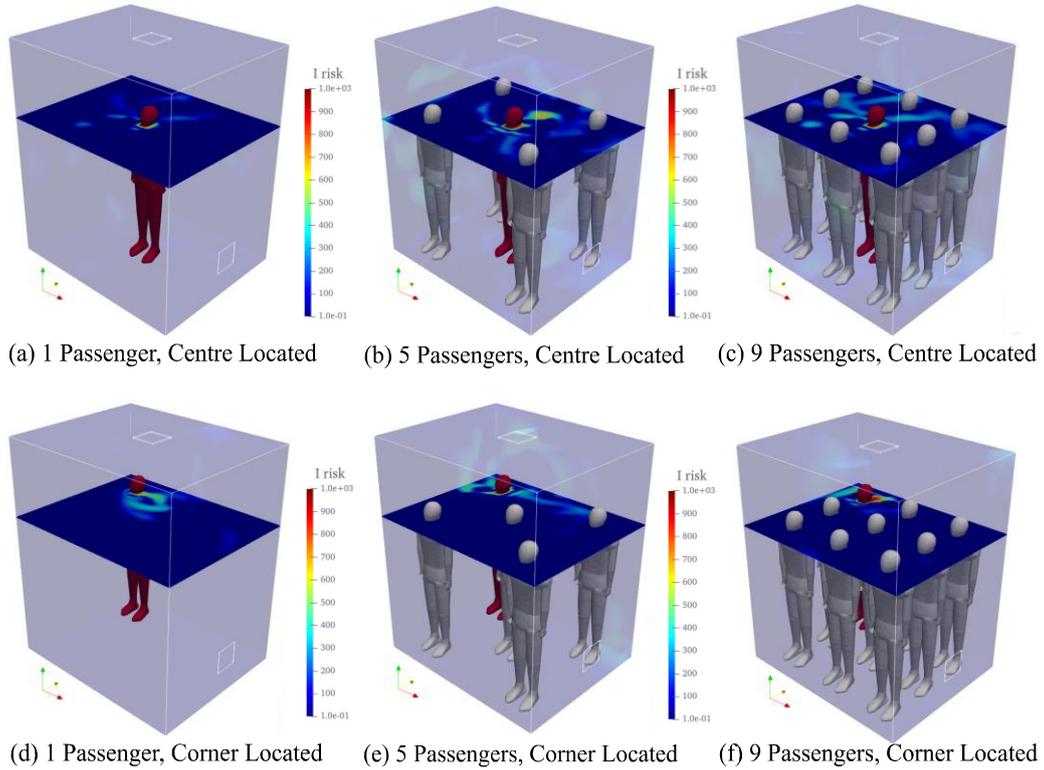

| (a) 1 Passenger, Centre Located | (b) 5 Passengers, Centre Located | (c) 9 Passengers, Centre Located |
| (d) 1 Passenger, Corner Located | (e) 5 Passengers, Corner Located | (f) 9 Passengers, Corner Located |

**Figure 13**. Influence of capacities on particle dispersion for DV-1

**Table 2.** Summary of calculated criteria for elevator capacity evaluation.

| Ventilation type | Source | Capacity | Suppression Effect (%) | Particle removal efficiency (%) | Probability of infection risk (%) over the sampling surface |
|---|---|---|---|---|---|
| MV-7 | Center | 1 | 05.10 % | 75.90 % | 06.10 % |
|  | Corner | 1 | 05.30 % | 79.40 % | 07.30 % |
|  | Corner | 5 | 12.41 % | 69.35 % | 11.71 % |
|  | Corner | 5 | 11.83 % | 68.54 % | 11.31 % |
|  | Center | 9 | 15.32 % | 65.85 % | 12.78 % |
|  | Corner | 9 | 15.48 % | 65.78 % | 13.10 % |
| DV-1 | Center | 1 | 13.20 % | 74.79% | 18.80 % |
|  | Corner | 1 | 18.40 % | 42.77% | 09.00 % |
|  | Corner | 5 | 26.87 % | 39.32 % | 11.02 % |
|  | Corner | 5 | 26.25 % | 40.14 % | 11.14 % |
|  | Center | 9 | 30.45 % | 35.45 % | 19.89 % |
|  | Corner | 9 | 32.47 % | 35.78 % | 22.47 % |

## 3.3. Influence of ventilation rate under different ventilation designs

This section investigates the impact of different ventilation rates of fresh air inside the elevator cabin on the aerosol spread for various ventilation designs. To quantify the effects of air change rates (ACHs), simulations were conducted at 2, 7, and 12 hr$^{-1}$, and particle removal efficiency, infection risk probability, and suppression effect were calculated for each scenario. **Figure 14** illustrates the infection risk distribution of MV-7 and DV-1 for both center and corner-located infectors at low (2 hr$^{-1}$), moderate (7 hr$^{-1}$), and high (12 hr$^{-1}$) ventilation rates. These results show the influence of ACH on the breathing aerosol spread inside the elevator at a temperature of 20 $^{o}$C. In poorly ventilated conditions, weak shear flow cannot overcome the aerosol cloud, and as a result, the zone with high particle concentration and aerosol index risk levels is created, leading to increased virus transmission. These zones are spread throughout the elevator cabin, allowing aerosol particles to move in any direction, consistent with the findings of Nazari et al. (2022) and Wang & Hong (2023).

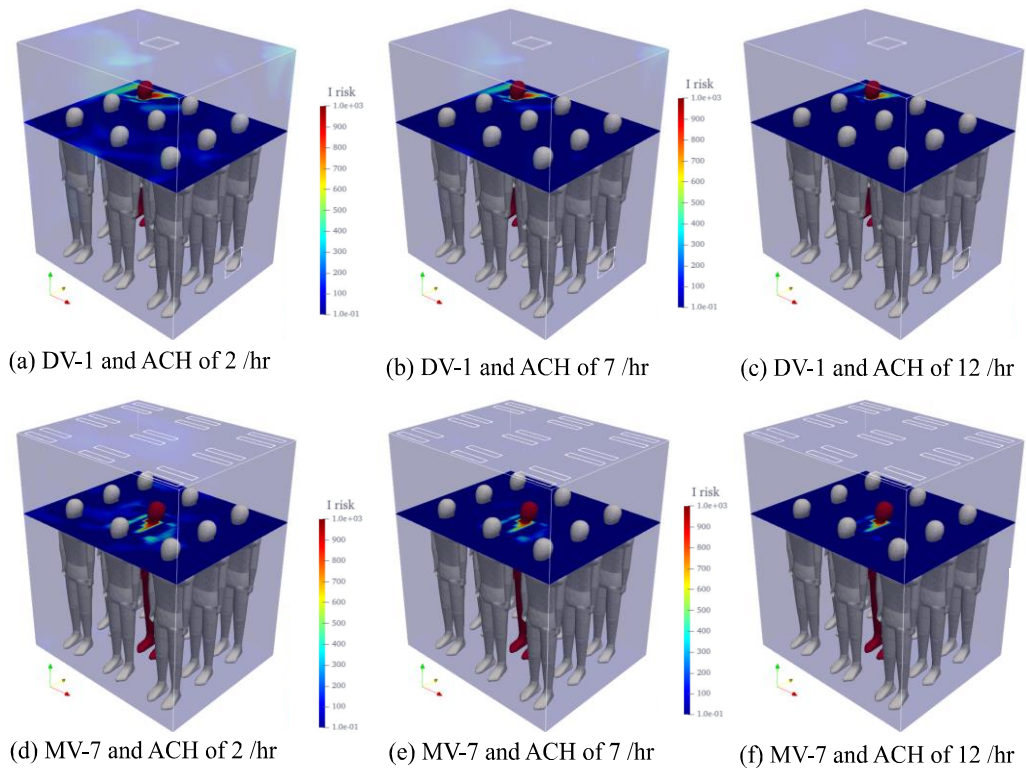

(a) DV-1 and ACH of 2 /hr    (b) DV-1 and ACH of 7 /hr    (c) DV-1 and ACH of 12 /hr

(d) MV-7 and ACH of 2 /hr    (e) MV-7 and ACH of 7 /hr    (f) MV-7 and ACH of 12 /hr

**Figure 14**. Influence of ACH on particle dispersion for MV-7 and DV-1.

**Table 3.** summary of calculated criterions in present work for elevator ACH studying of DV-1 and MV-7 in full elevator capacity

| Ventilation type | Capacity | ACH | Suppression Effect (%) | Particle removal efficiency (%) | Probability of infection risk (%) over the sampling surface |
|---|---|---|---|---|---|
| MV-7 | 9 (Centre) | 2 | 38.80 % | 40.23 % | 19.78 % |
|  | 9 (Centre) | 7 | 15.48 % | 65.78 % | 13.10 % |
|  | 9 (Centre) | 12 | 09.78 % | 69.78 % | 10.25 % |
| DV-1 | 9 (Corner) | 2 | 57.89 % | 30.78 % | 23.78 % |

| | | | | |
|---|---|---|---|---|
| 9 (Corner) | 7 | 30.45 % | 35.45 % | 19.89 % |
| 9 (Corner) | 12 | 10.89 % | 41.25 % | 11.25 % |

## 3.4. Comparison of the studied HVAC designs

The present study evaluated the designed elevator HVAC systems based on three key aspects, namely the suppression effect, particle removal efficiency, and infection risk. The results are summarized in **Table 4**, which lists the advantages and disadvantages of each design. The probability of infection risk represents the aerosol behavior on the sampling surface, while particle removal and suppression effect represent fluid flow behavior inside the elevator. The comparison of all designs from these three aspects provides a comprehensive overview for designers of various types of air conditioning systems, making this research an essential reference for future HVAC system designs.

**Table 4.** Summary of advantage and disadvantages of all designed elevator HVAC system.

| Ventilation type | Advantages | Disadvantages |
|---|---|---|
| MV-1 | Acceptable suppression effect. | Unacceptable particle removal efficiency and probability of infection risk, due to dispersion of aerosol by interaction of suppression effect of airflow and wall, capturing aerosol cloud not rapidly exhaust. |
| MV-2 | Due to streamline pattern, capturing aerosol cloud rapidly exhaust in the corner located case. | Unacceptable particle removal efficiency and probability of infection risk. |
| MV-3 | Similar airflow pattern to MV-2, due to streamline pattern, capturing aerosol cloud rapidly exhaust in the near of elevator wall. | Unacceptable particle removal efficiency and probability of infection risk. |
| MV-4 | Acceptable suppression effect, due to regular streamline, capturing aerosol cloud with down warding airflow and exhaust in the center located case. | Unacceptable particle removal efficiency and probability of infection risk, due to dispersion of aerosol by interaction of suppression effect of HVAC and non-slip wall, capturing aerosol cloud not rapidly exhaust. |
| MV-5 | - | Unacceptable particle removal efficiency and probability of infection risk, due to dispersion of aerosol by interaction of suppression effect of HVAC and non-slip wall, capturing aerosol cloud not rapidly exhaust. |
| MV-6 | Good probability of infection risk, due to streamline pattern, capturing aerosol cloud rapidly exhaust in the center located case. | Lower suppression effect that can increase probability of infection risk in failing situation of HVAC ducts. |
| MV-7 | Good particle removal and probability of infection risk, due to regular and separated streamline, capturing aerosol cloud rapidly exhaust in each position. | Lower suppression effect that can increase probability of infection risk in failing situation of HVAC ducts. |
| DV-1 | Acceptable particle removal and probability of infection risk, due to regular streamline, capturing aerosol cloud with up warding airflow. | Lower suppression effect that increases probability of infection risk in failing situation of HVAC ducts. |
| SV-1 | Good suppression effect | Unacceptable particle removal efficiency and infection risk probability. |
| SV-2 | Good suppression effect | Unacceptable particle removal efficiency and infection risk probability |

## 4. Conclusions and Discussions

In this study, we investigate the influence of various HVAC designs on the propagation of viruses (e.g., SARS-CoV-2) within elevator cabins. By utilizing a transport equation for aerosol concentration and validating our simulation approach with experimental data from Zhang et al. (2006), we investigate airborne transmission patterns and associated infection risks for different ventilation designs and source locations.

Our findings indicate that the choice of HVAC system plays a crucial role in mitigating the risk of airborne transmission of viruses in elevator cabins. The MV-7 system emerges as the most effective in reducing particle spread, achieving a maximum removal efficiency of 79.40% for a single passenger. The DV-1 system also performs well but exhibits slightly lower particle removal efficiency compared to the MV-7 system. In the center location, the DV-1 system demonstrates a removal efficiency of 74.79%, which declines to 42.77% in the corner location. Overall, the MV-7 system exhibits superior particle removal efficiency in both center and corner locations. Conversely, the stratum ventilation system (SV-2) presents the highest infection risk, with a particle removal efficiency of a mere 3.97%. Our investigation also reveals that infection risk is greater for corner-located sources compared to center-located sources, as the former generates more turbulence and increases air mixing, leading to heightened infection risk. Additionally, we determine that the infection risk within elevators rises with the number of occupants.

To reduce infection risk, we advocate for the implementation of HVAC systems that enhance both the ventilation rate and the operation time of elevator ventilation systems. An increased ventilation rate can minimize infection risk, while a longer operation time for the elevator ventilation system can further mitigate infection risk. We also encourage future research to examine the impact of elevator movement direction on aerosol concentration within cabins and to develop strategies for improving HVAC systems to manage viral transmission while maintaining comfortable conditions. Lastly, it is vital to account for energy consumption and thermal comfort when considering elevator HVAC systems in future studies.In conclusion, this study provides valuable insights into the mechanisms of virus spread via airborne transmission in elevator cabins and highlights the importance of carefully selecting and designing HVAC systems to reduce the risk of infection. Our findings have significant implications for public health and safety, particularly in the context of the COVID-19 pandemic. Our proposed DV-1 and MV-7 models are an effective solution to reduce the risk of airborne transmission in high-capacity density and confined spaces, such as elevator cabins. This study acknowledges several limitations that warrant further exploration. A primary constraint is the absence of consideration for aerosol dispersion generated by sneezing or coughing within the elevator cabin during the HVAC system design process. Expanding our understanding of virus transmission within the HVAC system under these conditions is essential for future research. Additionally, it is important to investigate the potential impact of varying acceleration speeds in elevators, as observed in different countries, on aerosol cloud behavior during movement. A deeper understanding of this factor will undoubtedly contribute to a more comprehensive knowledge of HVAC systems in confined spaces such as elevators. Despite these limitations, the current study's findings can still provide valuable insights for the development of more efficient and safer elevator HVAC systems, ultimately promoting public health and safety in enclosed environments.

**Acknowledgement**

C. Wang would like to thank the financial support of the China Scholarship Council (CSC, Grant No. 201906030144).

**Data Availability:** The data that support the findings of this study are available from the corresponding author upon reasonable request.

**Conflict of Interest:** The authors declare no conflict of interest.